\documentclass[12pt, border=55mm,preprintnumbers]{article}

\usepackage{amssymb}
\usepackage{amsmath}

\usepackage{graphicx}
\usepackage{color}
\usepackage{relsize}
\usepackage{lmodern}

\usepackage{float}

\def\bea{\begin{eqnarray}}
\def\eea{\end{eqnarray}}
\def\bean{\begin{eqnarray*}}
\def\eean{\end{eqnarray*}}

\def\bea{\begin{eqnarray}}
\def\eea{\end{eqnarray}}
\def\bean{\begin{equation*}}
\def\eean{\end{equation*}}

\begin{document}

\thispagestyle{empty}

\noindent\
\\
\\
\\
\begin{center}
\large \bf Dark Matter and Baryogenesis
from Non-Abelian Gauged Lepton Number\footnote{Plenary talk given at the Conference on Cosmology, Gravitational Waves and Particles, Singapore, February 6-10, 2017; based on the work: B.~Fornal, Y.~Shirman, T.\,M.\,P.~Tait and J.\,R. West,
arXiv:1703.00199 [hep-ph] \cite{Fornal:2017owa}.}
\end{center}
\hfill
 \vspace*{1cm}
\noindent
\begin{center}
{\bf Bartosz Fornal}\\ \vspace{2mm}
{\emph{Department of Physics, University of California, San Diego \\
9500 Gilman Drive, La Jolla, CA 92093, USA}}
\vspace*{1.5cm}
\end{center}

\begin{abstract}
A simple model is constructed based on the gauge symmetry  $SU(3)_c$ $\times \, SU(2)_L \times U(1)_Y \times SU(2)_\ell$, with only the leptons transforming nontrivially under $SU(2)_\ell$. The extended symmetry is broken down to the Standard Model gauge group at TeV-scale energies. We show that this model provides a mechanism for baryogenesis via leptogenesis in which the lepton number asymmetry is generated by $SU(2)_\ell$ instantons. The theory also contains a dark matter candidate -- the $SU(2)_\ell$ partner of the right-handed neutrino.
\end{abstract}

\newpage

\section{Introduction}\label{aba:sec1}
The Standard Model of elementary particle physics provides an extremely accurate description of Nature at the most fundamental level. Despite its remarkable successes, it explains only 5\% of the Universe, while the remaining 95\% is attributed to the mysterious  dark matter and dark energy.  In addition, the Standard Model itself has its own shortcomings: the inability to  generate the observable matter-antimatter asymmetry of the Universe, the hierarchy problem, massless neutrinos, unknown origin of flavor, and many more. Although a plethora of  models dealing with those issues have been constructed, it is still an open question which of those theories, if any, provides the correct or at least partially correct description of Nature at higher energies. We simply need more experimental data to find this out. In the meantime, further systematizing and rethinking of our model building efforts is definitely required.

The Standard Model is based on the gauge group $SU(3)_c \times SU(2)_L \times U(1)_Y$ \cite{Glashow:1961tr,Weinberg:1967tq,Salam:1968rm,SU(3),Fritzsch:1973pi}. Apart from this local symmetry, it also has two accidental global symmetries: baryon number and lepton number. One might wonder whether those are just residual symmetries left over from the breaking of a more fundamental extended gauge symmetry. Efforts of gauging baryon and lepton number were carried in the past \cite{Pais:1973mi,Tosa:1982pv,Rajpoot:1987yg,Foot:1989ts,Carone:1995pu,Georgi:1996ei}, but only the models constructed recently \cite{FileviezPerez:2010gw,Duerr:2013dza,Schwaller:2013hqa,Arnold:2013qja} are experimentally viable. In theories of this type gauge coupling unification does not occur naturally and so far only partial unification has been achieved \cite{Fornal:2015one,Fornal:2016boa}. 

Nevertheless, simple extensions of the Standard Model gauge group provide a good playground for testing various approaches to the dark matter and baryogenesis puzzles. In this talk, I will discuss one of such extensions containing a dark matter candidate and  offering a mechanism for producing a lepton asymmetry, which ultimately can explain the matter-antimatter asymmetry of the Universe.

{\renewcommand{\arraystretch}{1.4}\begin{table}[t!]
\begin{center}
    \begin{tabular}{| c || c || c | c |} \hline
      \  \ \ Field \ \ \  & \ \ $SU(2)_\ell$ \ \ & \ \ $SU(2)_L$ \  \ & \ \ $U(1)_Y$  \ \       \\ \hline\hline
         \ \ $\hat{l}_{L}  \ \ $ & $2$ & $2$ & $-\frac12$  \\  [2pt] \hline
         $\hat{e}_R $  & $2$ & $1$ &  $-1$  \\ \hline
          $\hat{\nu}_R $  &  $2$ & $1$ &  \ $0$   \\ \hline\hline
               $l^\prime_{R}$  & $1$ & $2$ & $-\frac12$ \\ [2pt] \hline
         $e'_L$  & $1$ & $1$ &  $-1$  \\ \hline
          $\nu'_L$ & $1$ & $1$ &  \ $0$  \\ \hline\hline
        $ \ {\hat\Phi}_{1,2}$ & $2$ & $1$ & \  $0$  \\ \hline
    \end{tabular}
\end{center}
\caption{\small{New field representations in the model.}}
\label{tab1}
\end{table}

\section{The model}
The theory we propose is based on the gauge group:
\bea
SU(3)_c \times SU(2)_{L} \times U(1)_Y  \times SU(2)_{\ell} \ .
\eea

\subsection{Fermionic sector}

The Standard Model quarks are singlets under $SU(2)_\ell$, whereas the left-handed lepton doublet $l_L$ and the right-handed electron $e_R$ are the upper components of $SU(2)_\ell$ doublets,
\bea
&& \hspace{-10mm}\hat{l}_L =  \left(\begin{array}{c}
    l\\
    \tilde{l} \\
  \end{array} \right)_{\!L}  , \ \ \ \ \hat{e}_{R} = \left(\begin{array}{c}
    e\\
    \tilde{e} \\
  \end{array} \right)_{\!R}  ,
\eea
where the lower components are the new partner fields $\tilde{l}_L$ and $\tilde{e}_R$, respectively. 
To cancel the gauge anomalies involving $SU(2)_\ell$ one requires an extra $SU(2)_\ell$ doublet of Standard Model singlet fields,
\bea  \hat{\nu}_{R} =  \left(\begin{array}{c}
    \nu\\
    \tilde{\nu} \\
  \end{array} \right)_{\!R} .
\eea
The remaining anomalies involving just the Standard Model gauge groups are canceled by introducing new $SU(2)_\ell$ singlet fields:
\bea
&& {l}'_{R}  , \ \ \ \ {e}'_{L}  , \ \ \ \ {\nu}'_{L} \ .
\eea
The particle content of the model along with the quantum numbers of the fields is shown in Table~\ref{tab1}. The Standard Model quarks were not included since they transform trivially under $SU(2)_\ell$.

\subsection{Higgs and gauge sector}
Although breaking of the extended gauge group down to the Standard Model can be achieved with just one new $SU(2)_\ell$ doublet Higgs, for reasons discussed later we introduce two new Higgs fields $\hat\Phi_{1,2}$ and assume that one of the vacuum expectation values (vevs) is much larger than the other, $v_1 \gg v_2$. This can be easily engineered by choosing appropriate values for the parameters  in the scalar potential:
\bea
V(\Phi_1,\Phi_2) &=&  m_1^2 |\hat\Phi_1|^2 + m_2^2 |\hat\Phi_2|^2+ (m_{12}^2 \hat\Phi_1^\dagger \hat\Phi_2 + \rm{h.c.}) \nonumber\\
&+& \lambda_1 |\hat\Phi_1|^4 +\lambda_2 |\hat\Phi_2|^4 +\,\lambda_3 |\hat\Phi_1|^2  |\hat\Phi_2|^2 + \lambda_4 |\hat\Phi_1^\dagger\hat\Phi_2|^2  \nonumber\\
 &+&   \Big[ \tilde{\lambda}_5 \hat\Phi_1^\dagger \hat\Phi_2 |\hat\Phi_1|^2 +\tilde{\lambda}_6 \hat\Phi_1^\dagger \hat\Phi_2 |\hat\Phi_2|^2 + \tilde{\lambda}_7 (\hat\Phi_1^\dagger \hat\Phi_2)^2 + \ {\rm h.c.}\Big],
 \label{V}
\eea
where we neglected terms involving the Standard Model Higgs field. 
After $\hat\Phi_{1,2}$ develop vevs,
\bea
\langle \hat\Phi_i \rangle = \frac{1}{\sqrt2} \left(\!
  \begin{array}{c}
    0\\
    v_i \\
  \end{array}\!
\right),
\eea
 the $SU(3)_c \times SU(2)_L \times U(1)_Y \times SU(2)_\ell$ symmetry is broken down directly to the Standard Model  gauge group. 
 
 Apart from the new Higgs particles, the theory contains three new vector gauge bosons: 
\bea
Z' \, ,  \ \ {W_+'}, \  \ {W_-'} \ ,
\eea
which do not mix with the Standard Model electroweak gauge bosons.

\subsection{Particle masses}
The Yukawa part of the Lagrangian is given by:
\bea
 \mathcal{L}_{\rm Y} &=&   \sum_i \left[Y_l^{ab}\,\bar{\hat{l}}_L^a \, \hat{\Phi}_i\,  {l'}_{\!\!R}^{b}+ Y_e^{ab} \, \bar{\hat{e}}_R^a \, \hat{\Phi}_i\, {e'}_{\!\!L}^{b}
   + Y_\nu^{ab}\,\bar{\hat{\nu}}_R^a \, \hat{\Phi}_i\, {\nu'}_{\!\!L}^{b}\right]\nonumber\\  \nonumber\\[-7pt]
 &+&   y_e^{ab}\, \bar{\hat{l}}_L^a \,H \,\hat{e}_R^b + y_\nu^{ab}\, \bar{\hat{l}}_L^a\, \tilde{H} \,\hat{\nu}_R^b + {y'}_{\!\!e}^{ab}\, \bar{l'}_{\!\!R}^a \,H \,{e'}_{\!\!L}^b + {{y'}}_{\!\!\nu}^{ab}\, \bar{l'}_{\!\!R}^a\,\tilde{H} \, {\nu'}_{\!\!L}^b + {\rm h.c.} \ ,
  \label{lagr}
\eea
where $a,b=1,2,3$ are flavor indices. After $SU(2)_\ell$ symmetry breaking, the Yukawa matrices $Y_l$, $Y_e$ and $Y_\nu$ lead to vector-like masses for all  new fermions in the theory. The Yukawa matrices $y_e$ and $y_\nu$ produce the usual Standard Model lepton masses and, along with $y_e'$ and $y_\nu'$, contribute also to lepton partner masses. The  fermionic mass matrix  is given by:
\bea
& & \hspace*{-0.75cm}
\tfrac{1}{\sqrt{2}}
\left( \, 
\overline{\tilde{\nu}}_{\!L} ~~ \overline{\nu}^\prime_L \, 
\right)
 \left(
  \begin{array}{cc}
    Y_l \,v_\ell & y_\nu v \\
    {y'}_{\!\!\nu}^\dagger v & Y_\nu^\dagger  v_\ell \\
  \end{array}
\right)
\left( \!
  \begin{array}{c}
   \nu^\prime_R \\
    \tilde{\nu}_R \\
  \end{array}\!
\right)+ \tfrac{1}{\sqrt{2}}
\left( \,
\overline{\tilde{e}}_L ~~ \overline{e}^\prime_L \,
\right)
 \left(
  \begin{array}{cc}
    Y_l \,v_\ell & y_e v \\
    {y'}_{\!\!e}^\dagger v &Y_e^\dagger  v_\ell \\
  \end{array}
\right)
\left(\!
  \begin{array}{c}
   e^\prime_R \\
    \tilde{e}_R \\
  \end{array}\!
\right)\nonumber\\
&&\hspace*{-0.5cm}+ \  {\rm h.c.} \ ,
\eea 
where $v_\ell = \sqrt{v_1^2+v_2^2}$ and $v$ is the Standard Model Higgs vev. The off-diagonal elements are due to the Yukawa terms involving the Standard Model Higgs and introduce mixing between the electroweak singlets and doublets. We assume $Y_{l, e, \nu} v_\ell \gg y_{e,\nu} v, y_{e,\nu}' v$, which is a phenomenologically natural assumption and frees the model from electroweak precision data constraints. 

The mass eigenstates consist of six electrically neutral and six electrically charged states. As shown below, after $SU(2)_\ell$ breaking there remains a residual global $U(1)_\ell$ symmetry which prevents the new particles from decaying to solely Standard Model states. Therefore, if the lightest of the mass eigenstates is electrically neutral, it becomes a natural candidate for dark matter. This implies that the dark matter particle in the model is the $SU(2)_\ell$ partner of the right-handed neutrino, which after electroweak symmetry breaking receives a small admixture from the electroweak doublets:
\begin{align}
\begin{aligned}
&\chi_L=\nu'_L + \epsilon \,\tilde\nu_L \ ,\\
&\chi_R=\tilde\nu_R + \epsilon \,\nu'_R \ ,
\label{chi}
\end{aligned}
\end{align}
where $\tilde\nu_L$ and $\nu_R'$ are the upper components of the electroweak doublets $\tilde{l}_L$ and $l_R'$, respectively, and $\epsilon\sim y_\nu v / (Y_\nu v_\ell)\sim y_\nu' v / (Y_\nu v_\ell)  \ll 1$. 

The Higgs spectrum of the theory is that of a generic  two Higgs doublet model. There are five physical scalar/pseudoscalar fields remaining after $SU(2)_\ell$ breaking. They are mixtures of the original CP-even and CP-odd components of $\hat\Phi_{1,2}$ and their masses depend on the choice of parameter values in the scalar potential (\ref{V}).

Regarding the gauge sector, since there is no mixing between $SU(2)_\ell$ and the other gauge groups, after $SU(2)_\ell$ breaking the new vector gauge bosons develop equal masses,
\bea
m_{Z', W_\pm'} \simeq \tfrac{1}{2} \,g_{\ell} \, v_\ell\ ,
\eea
where $g_\ell$ is the $SU(2)_\ell$ gauge coupling.

{\renewcommand{\arraystretch}{1.4}\begin{table}[t!]
\begin{center}
    \begin{tabular}{| c || c || c | c |} \hline
      \  \ \ Field \ \ \  & {$ \ \ U(1)_\ell \ \ $} & {$ \ \ U(1)_L \ \  $} & {$ \ \ U(1)_\chi \ \  $}        \\ \hline\hline
         \ \ $\hat{l}_{L}  \ \ $ & $1$ & $1$ & $0$   \\ \hline
         $\hat{e}_R $  & $1$ & $1$ & $0$  \\ \hline
          $\hat{\nu}_R $  & $1$ & $0$ & $1$   \\ \hline\hline
               $l^\prime_{R}$  & $1$ & $1$ & $0$ \\ \hline
         $e^\prime_L$  & $1$ & $1$ & $0$  \\ \hline
          $\nu^\prime_L$  & $1$ & $0$ & $1$  \\ \hline\hline
         $\ \Phi_{1,2}$  & $0$ & $0$ & $0$  \\ \hline
    \end{tabular}
\end{center}
\caption{\small{Charges under exact and approximate global symmetries.}}
\label{tab3}
\end{table}

\subsection{Global symmetries}
There exist two global symmetries of the Lagrangian. Only one of them, which we denote by $U(1)_\ell$, remains unbroken after $SU(2)_\ell$ breaking. Charges of the fields under this symmetry are provided in Table~\ref{tab3}. Using the fact that the Yukawa couplings $y_\nu$ are tiny to account for the smallness of the Standard Model neutrino masses, and assuming that $y_\nu'$ are small as well, the $U(1)_\ell$ global symmetry is promoted to two global symmetries, $U(1)_L$ and $U(1)_\chi$, which separately survive the breaking of $SU(2)_\ell$. The charges under those global symmetries are also shown in Table~\ref{tab3}. 
Note that the charge under $U(1)_L$ can be interpreted as Standard Model lepton number, whereas the $U(1)_\chi$ charge is the dark matter number.

\section{Baryogenesis}

We now discuss the details of the baryon number asymmetry generation in the model. It relies on the fact that a primordial lepton asymmetry is produced by $SU(2)_\ell$ instantons\footnote{We note that a similar idea was presented in Ref.~\cite{Blennow:2010qp}, which was brought to our attention after the completion of this work.}. The subsequent stages combine key features of several mechanisms: Dirac leptogenesis \cite{Dick:1999je,Murayama:2002je}, asymmetric dark matter \cite{Nussinov:1985xr,Kaplan:1991ah,Hooper:2004dc,Kaplan:2009ag,Petraki:2013wwa,Zurek:2013wia} and baryogenesis from an earlier phase transition \cite{Shu:2006mm}.

\subsection{$SU(2)_\ell$ instantons}
Because of the non-Abelian nature of $SU(2)_\ell$, the model exhibits  \, non-perturbative dynamics in the form of $SU(2)_\ell$ instantons, which are active only above the $SU(2)_\ell$ breaking scale. The instantons preserve the global $U(1)_\ell$ symmetry, but they do not conserve the global $U(1)_L$ and $U(1)_\chi$ symmetries  discussed in the previous section, since those symmetries are both anomalous under $SU(2)_\ell$ interactions. Following the calculation in Ref.~\cite{Morrissey:2005uza}, we find that the instantons induce the following dimension-six interaction terms:
\begin{align}
\begin{aligned}
& \epsilon_{ij} \ \Big[(l_L^i \cdot \bar{{\nu}}_R)(l_L^j \cdot \bar{{e}}_R) - (l_L^i \cdot \bar{{\nu}}_R)(\tilde{l}_L^j \cdot \bar{\tilde{e}}_R) + \ (l_L^i \cdot \tilde{l}_L^j)(\bar{\nu}_R \cdot \bar{\tilde{e}}_R)\\
&- \  (l_L^i \cdot \tilde{l}_L^j)(\bar{\tilde{\nu}}_R \cdot \bar{{e}}_R) +   (\tilde{l}_L^i \cdot \bar{\tilde{\nu}}_R)(\tilde{l}_L^j \cdot \bar{\tilde{e}}_R) - (\tilde{l}_L^i \cdot \bar{\tilde{\nu}}_R)({l}_L^j \cdot \bar{{e}}_R)\Big], \ \ \ \ \ 
\label{inst}
\end{aligned}
\end{align}
where the dots denote Lorentz contractions and, for simplicity, we assumed  just one generation of matter. The generalization to three families is straightforward.

\begin{figure}[t!]
  \centering
      \includegraphics[width=0.5\textwidth]{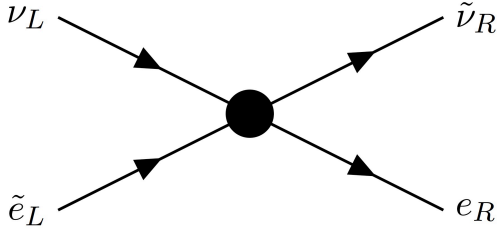}\vspace{-8mm}
  \caption{\small{One of the interactions induced by $SU(2)_\ell$ instantons for which $\Delta L = -1$ and $\Delta\chi=1$.}}\vspace{5mm}
  \label{fig:instanton}
\end{figure}

The last term in (\ref{inst}), for example, generates two interaction terms, one of which gives rise to $\nu_L \tilde{e}_L\rightarrow \tilde{\nu}_R \,e_R$ shown in Fig.~\ref{fig:instanton}. For this process, as can be read off from Table~\ref{tab3}, both the Standard Model lepton number and the dark matter number are violated by one unit:  $\Delta L = -1$ and $\Delta\chi = 1$, respectively. 
Therefore, the first condition for a successful leptogenesis, lepton number violation, is present in the model.

\begin{figure}[t!]
  \centering
      \includegraphics[width=0.7\textwidth]{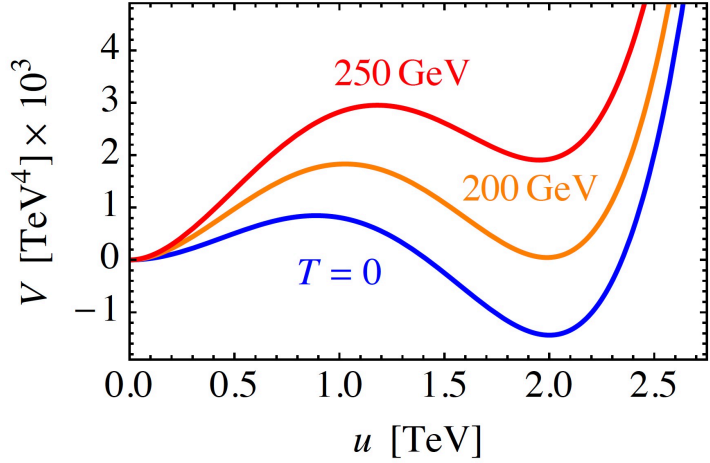}\vspace{-3mm}
  \caption{\small{Plot of the finite temperature effective potential  for $v_\ell = 2 \ \rm TeV$, $\lambda_{1} = 2\times 10^{-3}$ and $g_\ell = 1$.}}\vspace{3mm}
  \label{fig:finite_temp}
\end{figure}

\subsection{CP violation and phase transition}

The remaining Sakharov conditions \cite{Sakharov:1967dj} require sufficient CP violation and out-of-equillibrium dynamics, which in our model can be realized via a first order phase transition. 
The scalar potential (\ref{V}) contains four complex parameters: $m_{12}^2$, $\tilde\lambda_5$, $\tilde\lambda_6$, and $\tilde\lambda_7$. One phase can be rotated away by redefining the phase of $\hat\Phi^\dagger_1 \hat\Phi_2$, leaving three physical phase combinations \cite{Gunion:2005ja}. It is straightforward to show that for natural values of parameters the amount of CP violation in the model meets the criteria for a successful baryogenesis \cite{Fornal:2017owa}.

The last condition that needs to be checked is whether the model can actually accommodate a first order phase transition of the Universe. For this purpose we analyze the finite temperature effective potential \cite{Quiros:1999jp}, under the simplifying assumption $v_1 \gg v_2$:
\bea
V(u, T) \!\!\!&=&\!\!\!-\frac{1}{2} m_1^2 \,u^2 + \frac{1}{4} \lambda_1 \, u^4 \nonumber\\
&+& \!\!\! \frac{1}{64\pi^2} \sum_i n_i \left\{m_i^4(u)\left[\log\left(\frac{m_i^2(u)}{m_i^2(v_\ell)}\right)-\frac{3}{2}\right]+2\,m_i^2(u)\,m_i^2(v_\ell)\right\} \nonumber\\
&+& \!\!\!\frac{T^4}{4 \pi^2}\sum_i n_i \int_0^\infty \!d x \, x^2 \!\left[\log\!\left(1\mp e^{-\sqrt{x^2+ m_i^2(u)/T^2}}\right) - \log\!\left(1\mp e^{-x}\right)\right] \ \ \ \ \ \ 
\eea
where the first line is the tree-level Higgs contribution, the second line is the one loop zero temperature Coleman-Weinberg correction, and the last line is the finite temperature part. The sum is over all particles in the model, with appropriate factors corresponding to the number of degrees of freedom and statistics.  

The plot of the effective potential is shown in Fig.~\ref{fig:finite_temp} for a choice of parameters which sets the first order phase transition at the critical temperature $T_c = 200 \ {\rm{GeV}}$. We chose $v_\ell = 2 \ {\rm TeV}$, just above the current limit $v_\ell \gtrsim 1.7 \ {\rm TeV}$ set by the LEP-II experiment \cite{Schwaller:2013hqa}, so that the condition $v_\ell (T_c)/T_c \gtrsim 1$ for a strongly first order phase transition  is fulfilled.

\subsection{Bubble nucleation and lepton asymmetry}

\begin{figure}[t!]
  \centering
      \includegraphics[width=0.65\textwidth]{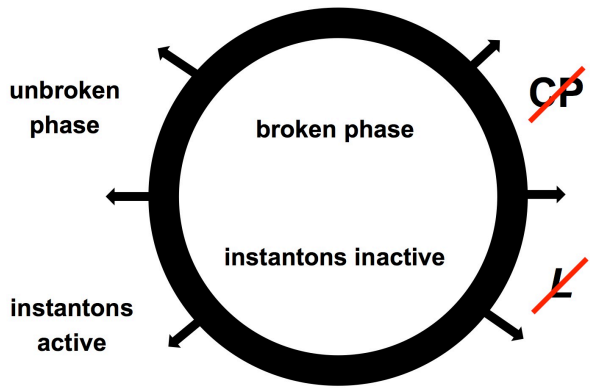}\vspace{-6mm}
  \caption{\small{Expanding bubble of true vacuum.}}\vspace{2mm}
  \label{fig:bubble}
\end{figure}

As the temperature of the Universe decreases and drops to $T_c$, bubbles of true vacuum start forming and expand, eventually filling out the entire Universe. A bubble expansion is schematically shown in Fig.~\ref{fig:bubble}. Outside the bubble the $SU(2)_\ell$ symmetry is not broken and the $SU(2)_\ell$ instantons remain active. Inside the bubble, on the other hand, $SU(2)_\ell$ is broken and the instanton effects are exponentially suppressed. As the bubble expands in the presence of CP violation, part of the lepton asymmetry generated by the instantons just outside the bubble becomes trapped inside the bubble. The same is true regarding the dark matter asymmetry. 

Although $SU(2)_\ell$ instantons are not active inside the bubble, one might worry whether the Standard Model lepton and dark matter asymmetries will be washed out by the Yukawa interactions involving $y_\nu$ and $y_\nu'$, since they explicitly violate $U(1)_L$ and $U(1)_\chi$, while conserving only their sum $U(1)_\ell$. This, however, is not an issue, since the small values of $y_\nu$ and $y_\nu'$ imply that the right-handed neutrinos and their partners reach chemical equillibrium long after the $SU(2)_\ell$ phase transition. As a result, the Standard Model lepton and dark matter number asymmetries survive until the electroweak phase transition, with just  the lepton asymmetry being partially converted into a baryon asymmetry by the electroweak sphalerons, as discussed in the subsequent subsection.

The process of accumulation of the Standard Model  lepton and dark matter asymmetries outside the expanding bubble is described by the diffusion equations \cite{Joyce:1994zn,Cohen:1994ss}:
\bea
\dot{n}_i = D_i \nabla^2 n_i - \sum_j \Gamma_{ij} \frac{n_j}{k_j} + \gamma_i \ ,
\eea
where $n_i$ denotes the number density of a given particle species, $D_i$ is the diffusion constant, $\Gamma_{ij}$ is the diffusion rate, $k_j$ is the number of degrees of freedom times a factor arising from statistics and $\gamma_i$ is the CP violating source \cite{Riotto:1995hh}. In our model there is a set of twelve diffusion equations and eight constraints coming from the Yukawa and instanton interactions \cite{Fornal:2017owa}.

\begin{figure}[t!]
  \centering
      \includegraphics[width=0.6\textwidth]{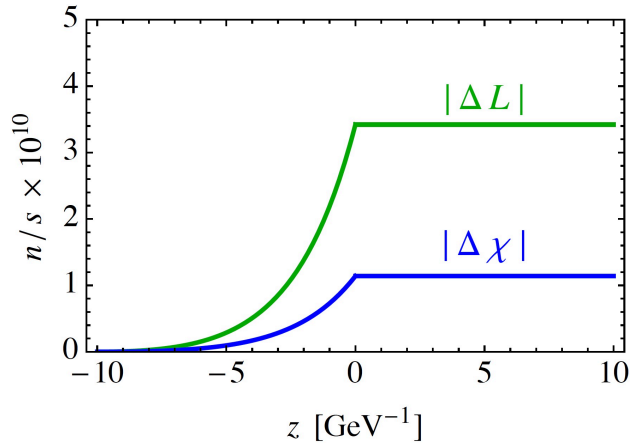}
  \caption{\small{Standard Model lepton and dark matter particle number densities vs. distance from the bubble wall for a set of natural parameter values.}}\vspace{5mm}
  \label{fig:diff_sol}
\end{figure}

The solution to this set of diffusion equations is shown in Fig.~\ref{fig:diff_sol}, which plots the  Standard Model lepton and dark matter particle number densities normalized to entropy as a function of the distance from the bubble wall located at $z=0$, where $z<0$ corresponds to the outside of the bubble and $z>0$ describes the inside of the bubble. 
The ratio of the generated Standard Model lepton and dark matter asymmetries in our model is:
\bea
\left|\frac{\Delta L}{\Delta \chi}\right| = 3 
\eea
and is independent of the model parameters. There exists a natural and experimentally allowed choice of parameters for which
\bea
\frac{n_L}{s} \simeq 3 \times 10^{-10} \ .
\eea

\subsection{Baryon asymmetry}
As mentioned earlier, the $SU(2)_\ell$ instantons become inactive after $SU(2)_\ell$ breaking and the dark matter number freezes in inside the bubble. This is not exactly the case for the Standard Model lepton asymmetry, since the Standard Model electroweak sphalerons remain active until the electroweak phase transition and convert part of the lepton number to baryon number. The resulting baryon asymmetry generated by the sphalerons is \cite{Harvey:1990qw}
\bea
\Delta B = \frac{28}{79} \,\Delta L  \ , 
\eea
therefore the final baryon asymmetry to entropy ratio is
\bea
\frac{n_B}{s} \approx 10^{-10} \ ,
\eea
which agrees with the observed value.

\section{Dark matter}

The dark matter candidate in our model (\ref{chi})  is composed mostly of the $SU(2)_\ell$ doublet partner of the right-handed neutrino. It is therefore  predominantly a Standard Model singlet, with only a small admixture of an electroweak doublet picked up by its interactions with the Standard Model Higgs field. Because the dark matter and baryon number asymmetries in our model are closely related (approximately equal at present time), the dark matter mass is uniquely determined by the observable ratio of the dark matter and baryonic relic densities.  Assuming the dark matter is relativistic at the decoupling temperature, its mass is given by:
\bea
m_\chi = m_p  \frac{\Omega_{\rm DM}}{\Omega_{\rm B}} \left|\frac{\Delta B}{\Delta \chi}\right| \simeq 5 \ {\rm GeV} \ .
\eea

A dark matter mass of a few GeV is generic in asymmetric dark matter models and it is generally challenging to make its symmetric component efficiently annihilate away. In the current model this issue is circumvented by arranging one of the Higgs components to be lighter than $5 \ {\rm GeV}$. In such a scenario a successful annihilation can proceed through the channels shown in Fig.~\ref{fig:Higgs_ann}. 
A light Higgs component can be realized provided that the quartic terms in the scalar potential are small, which in the case of $\lambda_1$ is also needed for a first order phase transition. This mass range for a new scalar/pesudoscalar is not strongly constrained by low energy experiments, but may be accessible in the future \cite{Krnjaic:2015mbs}.

\begin{figure}[t!]
  \centering
      \includegraphics[width=0.7\textwidth]{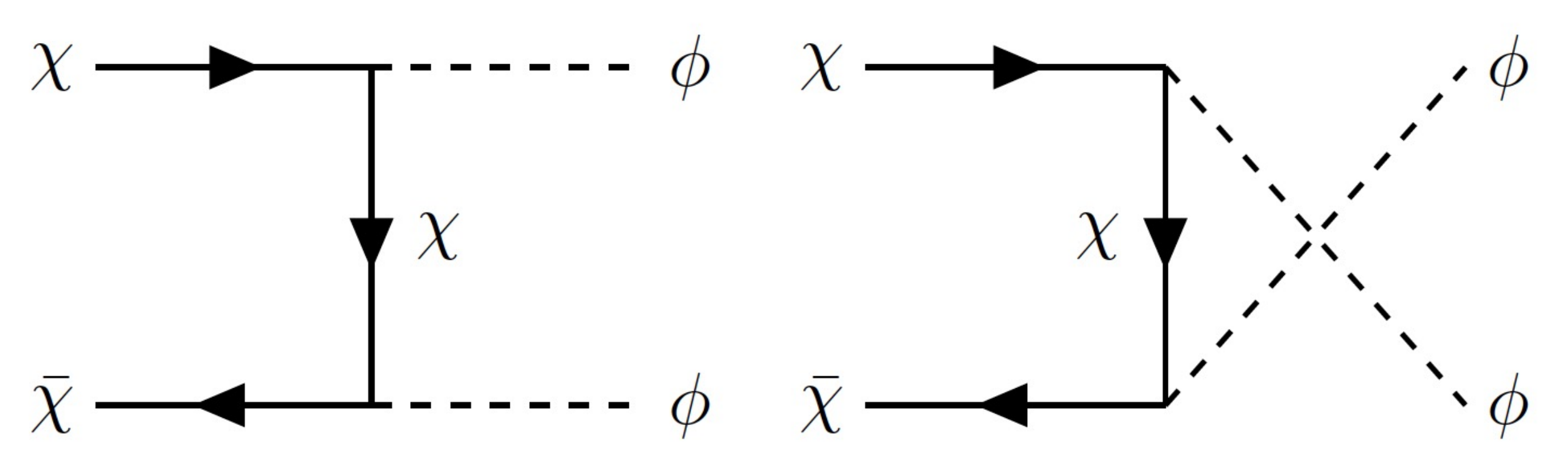}
  \caption{\small{Dark matter annihilation to the pseudoscalar component of $\hat\Phi_2$.}}
  \label{fig:Higgs_ann}\vspace{2mm}
\end{figure}

Since the new $Z'$ gauge bosons do not interact with quarks, there are no tree-level direct detection diagrams involving the Standard Model singlet component of the dark matter. As a result, the direct detection constraint in this case can only arise from loop processes, but  the GeV-scale dark matter limits coming from the CDMSlite experiment \cite{Agnese:2015nto} are much less restrictive than the LEP-II constraint of $v_\ell \gtrsim 1.7 \ {\rm TeV}$, which we already took into account. Regarding the contribution of the direct detection  diagrams involving the electroweak gauge bosons, it can be  estimated using the results of Ref.~\cite{Fornal:2016boa} and is fully consistent with experiment in the phenomenologically natural limit  $Y_{\nu} v_\ell \gg y_{\nu} v, y_{\nu}' v$ we adopted.

\section{Conclusions}

In this talk, I discussed a new model extending the Standard Model gauge group with a non-Abelian gauged lepton number $SU(2)_\ell$.  The model realizes a mechanism for baryogenesis based on leptogenesis in which the lepton number asymmetry is generated by $SU(2)_\ell$ instantons. It also contains a natural dark matter candidate --  the partner of the right-handed neutrino.
Despite the theoretical advantages, it is difficult to test this theory experimentally. Since new physics resides in the lepton sector of the model, the best way to probe it would be in a new high energy lepton-lepton collider. 

Let me end by saying that there is no reason not to expect the Standard Model gauge symmetry to be enhanced at higher energies. 
Analyzing other simple theories with extended gauge groups seems like a worthwhile effort, as it may shed more light on the outstanding issues of the Standard Model. However, one should always keep in mind that
``Nature will do what Nature does and it's up to experiment to be the final judge''\footnote{From: M.\,B.\,Wise, interviewed by A.\,Ananthaswamy, article title: \emph{Hunt is on for 
quark dark matter}, New Scientist, issue 3050, p.\,10, published on December 5,  2015.}.

\section*{Acknowledgments}
I am grateful to Yuri Shirman, Tim Tait and Jennifer West for a fruitful collaboration. 
 I would also like to thank the organizers of the Conference on Cosmology, 
 Gravitational Waves and Particles in Singapore, especially the chairman, Harald
Fritzsch, for the invitation, warm hospitality, and a fantastic scientific and
social atmosphere during the conference. This research was supported in
part by the DOE
grant DE-SC0009919 and the NSF grant PHY-1316792.

\bibliographystyle{utphys}
\bibliography{Nonabelian_lepton}

\end{document}